\documentclass{Interspeech}

\usepackage{cite}
\usepackage{amsmath,amssymb,amsfonts}
\usepackage{algorithmic}
\usepackage{graphicx}
\usepackage{textcomp}
\usepackage{xcolor}
\usepackage{multirow}
\usepackage{makecell}
\usepackage{amsfonts}

\interspeechcameraready

\title{SaD: A Scenario-Aware Discriminator for Speech Enhancement}

\author[affiliation={1}]{Xihao}{Yuan$^*$}
\author[affiliation={1}]{Siqi}{Liu$^*$}
\author[affiliation={1}]{Yan}{Chen}
\author[affiliation={1}]{Hang}{Zhou}
\author[affiliation={2}]{Chang}{Liu}
\author[affiliation={1}]{Hanting}{Chen}
\author[affiliation={1}]{Jie}{Hu}

\affiliation{Noah's Ark Lab}{Huawei}{China}
\affiliation{Consumer Business Group}{Huawei}{China}
\email{\{yuanxihao,liusiqi37,chenyan176,zhouhang25,liuchang3,chenhanting,hujie23\}@huawei.com}

\usepackage{comment}

\begin{document}

\maketitle
\def\thefootnote{*}\footnotetext{These authors contributed equally to this work}

\begin{abstract}
    
    Generative adversarial network-based models have shown remarkable performance in the field of speech enhancement. However, the current optimization strategies for these models predominantly focus on refining the architecture of the generator or enhancing the quality evaluation metrics of the discriminator. This approach often overlooks the rich contextual information inherent in diverse scenarios. In this paper, we propose a scenario-aware discriminator that captures scene-specific features and performs frequency-domain division, thereby enabling a more accurate quality assessment of the enhanced speech generated by the generator. We conducted comprehensive experiments on three representative models using two publicly available datasets. The results demonstrate that our method can effectively adapt to various generator architectures without altering their structure, thereby unlocking further performance gains in speech enhancement across different scenarios.
\end{abstract}

\section{Introduction}

Speech enhancement (SE) is of paramount importance in modern communication systems and has garnered significant attention due to its applications in various fields such as telecommunications, hearing aids, and speech recognition. The advent of deep learning has revolutionized SE, with deep neural network (DNN)-based approaches \cite{hu2020dccrn, tan2018convolutional, defossez2019demucs, bulut2020low} consistently demonstrating superior performance compared to traditional signal-processing-based methods \cite{boll1979suppression, lim1978all}.

A notable milestone in this domain was achieved with the introduction of SEGAN \cite{pascual2017segan}, which pioneered the application of generative adversarial network (GAN) to SE tasks and revealed their potential for further enhancing model performance. Since then, an increasing number of studies \cite{fu2019metricgan, su2020hifi, cao2022cmgan, close2024multi, zadorozhnyy2022scp, babaev2025finally} have focused on investigating and optimizing GAN-based models for SE, highlighting the growing interest in leveraging adversarial training to address the challenges of speech enhancement.

The GAN-like model comprises two core components: the generator and the discriminator, with the latter being crucial for evaluating the quality of generated results and guiding the generator to produce high-quality outputs. SEGAN employs a discriminator that merely distinguishes whether the generated result is a noisy signal or a clean speech signal, thereby neglecting the quality of the generated result and leading to suboptimal performance. MetricGAN \cite{fu2019metricgan} further introduces a metric-based discriminator that evaluates the perceptual evaluation of speech quality (PESQ) \cite{rix2001perceptual} and short-time objective intelligibility (STOI) \cite{taal2011algorithm}, significantly improving performance and achieving state-of-the-art (SOTA) results in SE tasks at the time. Inspired by MetricGAN, subsequent works such as MetricGAN+ \cite{fu2021metricgan+}, CMGAN \cite{cao2022cmgan}, and Multi-CMGAN \cite{close2024multi} have optimized the metric evaluation aspect within the discriminator to further enhance the performance of GAN-based models in SE tasks.

\begin{figure}[t]
  \begin{center}
  \includegraphics[width=\linewidth]{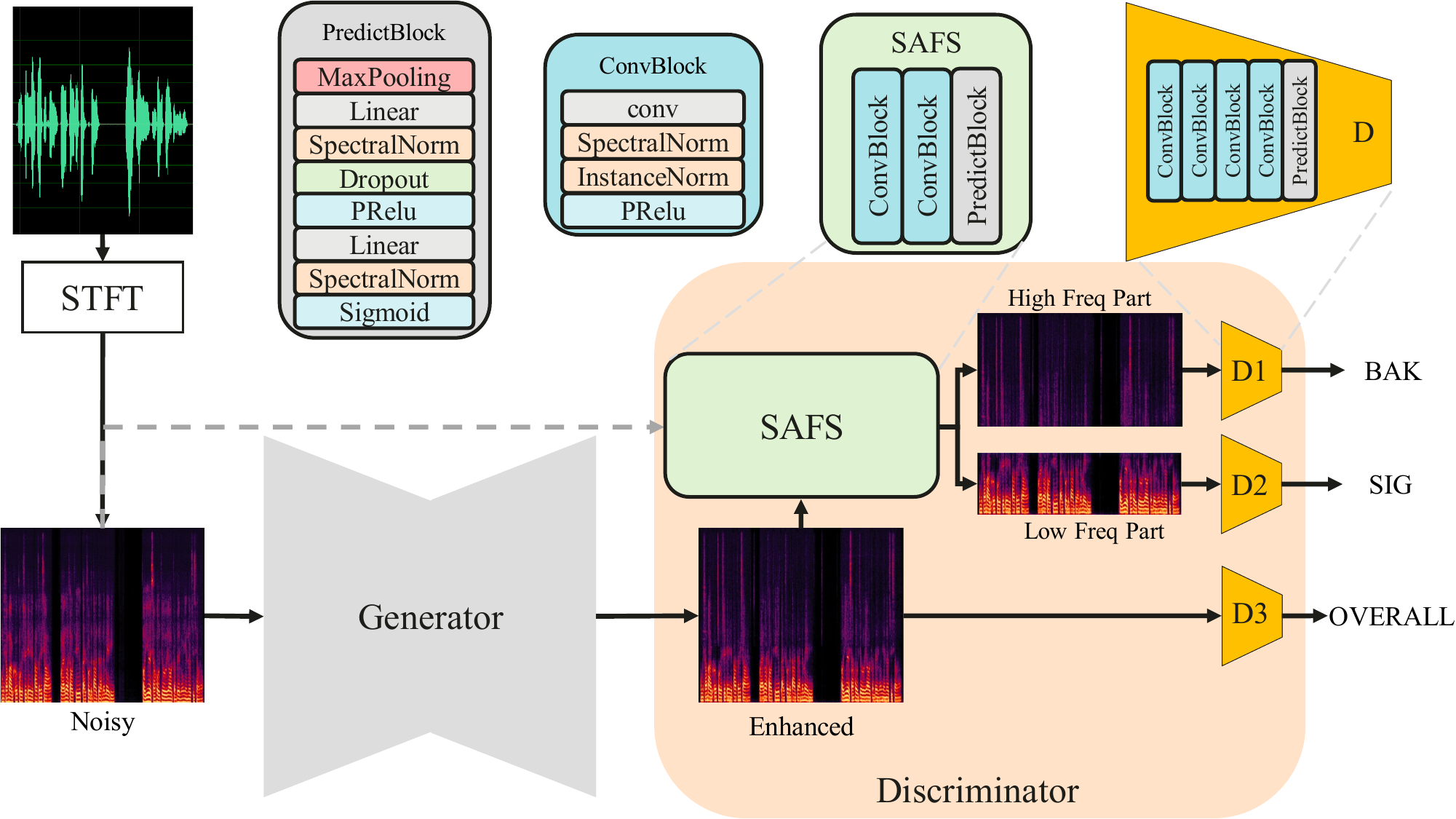}
  \caption{Overview of SaD. The Scenario-Aware Frequency Splitter receives the enhanced speech generated by the Generator and the original noisy speech as inputs, and predicts the frequency division points to partition the enhanced speech into high-frequency and low-frequency components. Three distinct pre-trained metric estimation discriminators are employed to evaluate the quality of the high-frequency component, low-frequency component, and the original enhanced speech, respectively.}
  \label{fig:SaD_overview}
  \end{center}
\end{figure}

In practical SE scenarios, environmental noise and human speech exhibit diverse distributions, resulting in varying frequency and signal-to-noise ratio (SNR) profiles. Despite extensive research on SE GAN models, existing efforts have predominantly focused on optimizing the discriminator through metric-based computations, with limited consideration of real-world scenario information. These discriminators often overlook the actual frequency distribution characteristics of different scenarios when evaluating quality. For instance, the frequency distribution of human speech is typically concentrated within the 1-4 kHz range, with speech being the dominant component below 4 kHz (low-frequency portion) and noise being the dominant component above 4 kHz (high-frequency portion). However, the actual frequency range of human speech varies among individuals (e.g., between male and female voices), implying that the precise frequency division point between high and low frequencies is not always strictly 4 kHz. Additionally, the frequency distribution of noise and the SNR vary across different scenarios, leading to differing proportions of speech and noise in various frequency bands.

To address this challenge, several existing methods have been proposed to enhance model performance by partitioning the frequency spectrum and fine-tuning individual frequency bands. For instance, Suband-KD \cite{hao2020sub} employs a fixed frequency band division strategy, training distinct teacher models for each band and subsequently distilling knowledge into a target model. However, this approach does not account for the actual characteristics of the acoustic scene. Another method, DFKD \cite{yuan2025dynamic}, estimates the frequency band division point by identifying the first-order derivative change extreme points in the frequency domain. While this technique incorporates scene characteristics to some extent, it relies heavily on empirical computation, which may not always yield optimal results. Therefore, fully integrating scene-specific characteristics and conducting a fine-grained evaluation of the denoising quality across different frequency bands is of paramount importance for enhancing the model's noise reduction capabilities.

In this paper, we introduce a scenario-aware discriminator (SaD) for speech enhancement to achieve finer differentiation and quality assessment of noise reduction across various acoustic scenarios. Drawing inspiration from the DFKD method, we propose a frequency band division approach based on weakly-supervised learning. This method enables the model to integrate scene-specific characteristics and adaptively generate frequency band divisions. Subsequently, distinct quality evaluation metrics are applied according to the signal characteristics of each band. By focusing more granularly on the signal distribution characteristics within different frequency bands, the model's performance is further enhanced. Our approach has been validated across multiple datasets and models, demonstrating significant improvements in both noise reduction and speech retention capabilities.

The remainder of this paper is structured as follows. Section 2 provides a detailed overview of the proposed method, highlighting its complexities. Section 3 delves into the specifics of the experiments, covering the dataset, implementation, and analysis of the results. Finally, Section 5 presents concluding remarks and outlines potential directions for future research.

\section{Methodology}

\subsection{System Overview}

Our proposed method reconfigures the discriminator for SE GAN-like models (e.g., MetricGAN, CMGAN). Specifically, we introduce a scenario-aware frequency splitter (SAFS) that adaptively partitions the enhanced speech generated by the generator into high-frequency and low-frequency components. The quality of these two frequency components is subsequently evaluated separately. Our approach is compatible with various generators of different architectures and has been validated as effective across multiple GAN-like models.

As illustrated in Fig. 1, the generator is assumed to accept time-frequency domain noisy speech processed by the short-time fourier transform (STFT) and transform it into enhanced speech. The SAFS in the proposed SaD module fuses the original noisy speech with the enhanced speech generated by the generator and predicts a division point. Based on this division point, the enhanced speech is divided into two parts: the low-frequency part, where speech is the dominant component, and the high-frequency part, where noise is the dominant component. Subsequently, distinct discriminators evaluate the quality of the signals in the high-frequency and low-frequency parts, as well as in the full-frequency band, and update the network parameters accordingly. Inspired by CMGAN, we combine the ConvBlock and PredictBlock to form the SAFS module and discriminator module, respectively.

\subsection{Scenario-Aware Frequency Spliter}

Another critical challenge is the absence of frequency division labels in existing SE datasets. Our experimental findings indicate that a fully unsupervised training approach is highly susceptible to pattern collapse. To mitigate this issue, we propose a weakly supervised training method leveraging the first-order derivative change extreme points. Specifically, we adopt the method of DFKD to calculate the frequency division point of the clean speech as the label for frequency division and utilize this label to guide the initial training phase of the SaD-GAN network. As the network converges, we remove the frequency division label and switch to an unsupervised training process. This transition allows the network to adaptively identify the optimal crossover frequency point in the later stages of training, in conjunction with the generator's actual speech enhancement capabilities and the characteristics of the current scenario.

Let $\mathcal{F}$ represent the SAFS, which takes as input the original noisy speech $X$ and the enhanced speech $\hat{Y}$ generated by the generator. The SAFS predicts the frequency division point $\hat{m}$. The ground-truth label $m$ is estimated using the methodology of DFKD. In the early stages of training, the loss function $Loss_{m}$ is employed to supervise the SAFS.

\begin{equation}
    \hat{m} = \mathcal{F}(X, \hat{Y})
\end{equation}

\begin{equation}
\begin{aligned}
    \hat{Y}_{high} = \hat{Y}[:\hat{m}], \hat{Y}_{low} = \hat{Y}[\hat{m}:]
\end{aligned}
\end{equation}

\begin{equation}
    Loss_{m} = \Vert m - \hat{m} \Vert_2
\end{equation}

\subsection{Quality Assessment of Different Frequencies}

The quality assessment method of DNSMOS \cite{reddy2021dnsmos} aligns more closely with human auditory perception. Recent studies leveraging DNSMOS have demonstrated its efficacy in discriminating quality and achieving satisfactory performance in SE tasks. However, this approach is not directly applicable to scenarios involving frequency division. Specifically, when the SaD partitions enhanced speech, the uncertainty of the division point introduces a dynamic signal, where the lengths of the high- and low-frequency bands are not fixed. Conversely, the conventional DNSMOS framework accepts only full-band inputs and evaluates three key metrics: background intrusiveness (BAK), speech distortion (SIG), and overall quality (OVERALL). This method is insufficient for accurately assessing the quality of individual high- or low-frequency components. To address this limitation and accommodate SaD, we retrained the discriminator and developed a system for evaluating the quality of high- and low-frequency separation.

To collect dynamic inputs for high- and low-frequency components, we adopt the approach of DFKD to partition the original clean speech from the training set into two frequency bands: high-frequency and low-frequency parts. Subsequently, we input the original clean speech into the pre-trained DNSMOS model to obtain the BAK, SIG, and OVERALL scores. The high-frequency part is then fed into the discriminator $\mathcal{D}_1$, which is supervised by the BAK score, while the low-frequency part is fed into the discriminator $\mathcal{D}_2$, supervised by the SIG score. Additionally, the enhanced speech is fed into the discriminator $\mathcal{D}_3$, which is supervised by the OVERALL score to ensure overall quality control. In this manner, $\mathcal{D}_1$ and $\mathcal{D}_2$ receive dynamic input signals and accurately estimate the noise suppression ability in the high-frequency part and the vocal retention ability in the low-frequency part, respectively.

\begin{equation}
    Loss_{BAK} = \Vert \mathcal{D}_1(\hat{Y}_{high}) - BAK \Vert_2
\end{equation}

\begin{equation}
    Loss_{SIG} = \Vert \mathcal{D}_2(\hat{Y}_{low}) - SIG \Vert_2
\end{equation}

\begin{equation}
    Loss_{OVL} = \Vert \mathcal{D}_3(\hat{Y}) - OVERALL \Vert_2
\end{equation}

Furthermore, the SNR information is crucial in SE tasks. However, previous SE GAN studies have often overlooked this aspect, with training processes typically calculating the discriminator loss ($Loss_\mathcal{D}$) solely based on final predicted metrics such as PESQ and STOI. This approach often necessitates complex parameter tuning to balance losses corresponding to different sub-metrics, thereby optimizing the network's overall performance.

By further analyzing the optimization goals of SE tasks within the context of scene SNR, it becomes evident that for high-SNR scenarios with minimal background noise, the primary objective of the network should be to preserve speech quality. Conversely, in low-SNR scenarios with significant background noise, noise suppression becomes more critical for enhancing speech intelligibility.

Based on these observations, we propose an SNR-driven loss balancing method. Here, the entire network is trained such that the SaD module adaptively adjusts the weights of different loss sub-items according to the SNR value of the current scenario. Specifically, $SNR_{max}$ denotes the maximum SNR value over the whole training dataset, the weight of $Loss_{BAK}$ is increased for low-SNR scenes, while the weight of $Loss_{SIG}$ is increased for high-SNR scenes. This adaptive weighting ensures that the network prioritizes noise suppression in challenging scenarios and speech preservation in simpler scenarios.

\begin{equation}
\begin{aligned}
    Loss_\mathcal{D} = & Loss_{m} + Loss_{OVL} + \\
    & \alpha * Loss_{BAK} + (1 - \alpha) * Loss_{SIG} \\
    & \text{where}\ \alpha = \frac{SNR}{SNR_{max}} \\
\end{aligned}
\end{equation}

\begin{equation}
    Loss_{total} = Loss_\mathcal{G} + \gamma * Loss_\mathcal{D}
\end{equation}

Finally, the total loss ($Loss_{total}$) during the training of the SaD-GAN model is obtained by combining the generator loss ($Loss_\mathcal{G}$) and the discriminator loss ($Loss_\mathcal{D}$), with their respective contributions regulated by the hyperparameter $\gamma$. It is important to note that our proposed method does not alter the generator architecture of the SE GAN model. Consequently, $Loss_\mathcal{G}$ remains strictly consistent with the original formulations in the literature and may vary across different implementations.

\section{Experiments}

\subsection{Datasets}

To validate the effectiveness of our proposed methods, we conducted experiments using the DNS2020 challenge dataset \cite{reddy2020interspeech} and the VoiceBank+DEMAND dataset \cite{valentini2016investigating}.

The dataset employed in the DNS2020 challenge comprises 500 hours of pristine speech recordings from 2,150 unique speakers, augmented with 65,000 noise clips representing 150 distinct audio classes. These noise clips were meticulously sourced from publicly available datasets, including Audioset, Freesound, and YouTube. To facilitate the training process, each audio clip was uniformly segmented into fixed intervals of 6 seconds. Subsequently, the entire training corpus was resampled at a consistent sampling rate of 16 kHz to ensure uniformity across all data points. Additionally, the SNR levels were randomly sampled from a uniform distribution ranging between 0 and 20 dB, thereby introducing variability to simulate diverse acoustic environments.

The VoiceBank+DEMAND dataset includes a training set of 11,572 recordings from 28 speakers, mixed with background noise from the DEMAND \cite{thiemann2013diverse} database and artificial sources at SNRs of 0, 5, 10, and 15 dB. The test set consists of 824 utterances from two speakers, combined with unseen DEMAND noise at SNRs of 2.5, 7.5, 12.5, and 17.5 dB.

\begin{table}[!tbp]
\caption{Performance comparison on VoiceBANK+DEMAND. \dag denotes the results tested by official open-source code, otherwise reproduced by ourselves.}
\begin{center}
\resizebox{7.2cm}{!}{
\centering

\begin{tabular}{cccccc}
\hline
\multirow{2}*{\textbf{Model}} & \multicolumn{5}{c}{\textbf{VoiceBANK+DEMAND test}} \\
\multicolumn{1}{c}{} & \textbf{PESQ} & \textbf{CSIG} & \textbf{CBAK} & \textbf{COVL} & \textbf{STOI} \\

\hline

\makecell{MetricGAN} & 2.697 & 3.861 & 2.428 & 3.291 & \textbf{0.876} \\
\makecell{MetricGAN + SaD} & \textbf{2.928} & \textbf{3.927} & \textbf{2.51} & \textbf{3.439} & 0.868 \\

\hline

\makecell{CMGAN\dag} & 3.406 & 4.595 & 2.831 & 4.076 & \textbf{0.958} \\
\makecell{CMGAN + SaD} & \textbf{3.622} & \textbf{4.659} & \textbf{3.24} & \textbf{4.223} & 0.947 \\

\hline

\makecell{Multi-CMGAN} & 3.393 & 4.421 & 3.349 & 3.953 & 0.942 \\
\makecell{Multi-CMGAN + SaD} & \textbf{3.402} & \textbf{4.465} & \textbf{3.372} & \textbf{3.982} & \textbf{0.943} \\

\hline

\end{tabular}
}
\end{center}
\end{table}

\begin{table}[!tbp]
\caption{Performance comparison on DNS2020.}
\begin{center}
\resizebox{7.2cm}{!}{
\centering

\begin{tabular}{cccccc}
\hline
\multirow{2}*{\textbf{Model}} & \multicolumn{5}{c}{\textbf{DNS2020-test}} \\
\multicolumn{1}{c}{} & \textbf{PESQ} & \textbf{CSIG} & \textbf{CBAK} & \textbf{COVL} & \textbf{STOI} \\

\hline

\makecell{MetricGAN} & 2.647 & \textbf{3.903} & \textbf{2.516} & 3.303 & \textbf{0.912} \\
\makecell{MetricGAN + SaD} & \textbf{2.663} & 3.88 & 2.457 & \textbf{3.304} & 0.894 \\

\hline

\makecell{CMGAN} & 3.107 & 4.283 & 3.156 & 3.717 & 0.926 \\
\makecell{CMGAN + SaD} & \textbf{3.135} & \textbf{4.445} & \textbf{3.197} & \textbf{3.833} & \textbf{0.936} \\

\hline

\makecell{Multi-CMGAN} & 3.013 & 4.353 & 3.186 & 3.717 & 0.944 \\
\makecell{Multi-CMGAN + SaD} & \textbf{3.115} & \textbf{4.378} & \textbf{3.239} & \textbf{3.779} & \textbf{0.948} \\

\hline

\end{tabular}
}
\end{center}
\end{table}

\subsection{Implementation Details}

To validate the generalizability of the proposed SaD module, we conducted experiments on several state-of-the-art SE GAN models, including MetricGAN, CMGAN, and the latest SOTA model, MultiCMGAN. These models were trained and tested on two above-mentioned benchmark datasets: DNS2020 and VoiceBank+DEMAND. 

During the training process, the discriminator in each model was replaced with the SaD module, while the generator architecture remained unchanged. For MetricGAN, the BLSTM architecture was employed as the generator, whereas for CMGAN and MultiCMGAN, the conformer architecture was utilized.

In the initial 10 epochs of SaD training, we employed the DFKD method to compute the division point labels, providing supervised guidance for the training process. Subsequently, the supervision was removed, enabling the network to adaptively learn the division points in an unsupervised manner. This approach ensures that the SaD module can dynamically adjust to varying acoustic environments while maintaining robust generalization capabilities.

For models with publicly available weights, we directly utilized these weights to evaluate their performance on the corresponding test sets. For models whose weights are not yet open-source, we meticulously adhered to the descriptions and hyperparameter settings outlined in the original literature, and trained and tested these models on the respective datasets.

In terms of evaluation metrics, we measured the PESQ, STOI, and the mean opinion score (MOS) \cite{hu2007evaluation} predictor, which includes three sub-metrics: speech distortion (CSIG), background noise intrusiveness (CBAK), and overall speech quality (COVL). Specifically, CSIG, CBAK, and COVL assess the signal distortion, background interference, and overall quality on a common scale, respectively. PESQ and STOI quantify the perceived quality and intelligibility of the speech signal, respectively.

\begin{table}[!tbp]
\caption{Ablation studies with CMGAN on VoiceBANK+DEMAND.}
\begin{center}
\resizebox{8cm}{!}{
\centering

\begin{tabular}{cccccc}
\hline
\multirow{2}*{\textbf{Model}} & \multicolumn{5}{c}{\textbf{VoiceBANK+DEMAND test}} \\
\multicolumn{1}{c}{} & \textbf{PESQ} & \textbf{CSIG} & \textbf{CBAK} & \textbf{COVL} & \textbf{STOI} \\

\hline

\makecell{\textbf{CMGAN + SaD}} & \textbf{3.622} & \textbf{4.659} & \textbf{3.24} & \textbf{4.223} & 0.947 \\
\makecell{\ \ \ \ \ \ \ \ \ \ \ \ \ w/o weakly supervised} & 3.539 & 4.646 & 3.103 & 4.158 & 0.849 \\
\makecell{\ \ \ \ \ \ \ \ \ \ \ \ \ \ \ w/o DNSMOS fine-tune} & 3.449 & 4.553 & 3.328 & 4.058 & 0.94 \\
\makecell{\ \ \ \ \ \ \ \ \ \ w/o SNR loss weight} & 3.59 & 4.636 & 3.139 & 4.182 & \textbf{0.951} \\

\hline

\end{tabular}
}
\end{center}
\end{table}

\subsection{Experimental Results and Discussions}

\subsubsection{Main Results}

As presented in Table 1, our proposed method was evaluated on three distinct models using the VoiceBank+DEMAND dataset. When the generator was kept consistent and only the discriminator was replaced with the proposed SaD module, our method consistently outperformed the original models across nearly all metrics. Notably, the PESQ scores for MetricGAN and CMGAN were enhanced by 0.2 points. Significant improvements were also observed in speech distortion, background intrusiveness, and overall quality. Among these models, CMGAN exhibited the most pronounced enhancement, with an increase of over 0.2 points in the overall quality score.

To further validate the effectiveness of our approach, additional experiments were conducted on the DNS2020 dataset, as detailed in Table 2. Our method demonstrated consistent improvements across all three models. Specifically, CMGAN and MultiCMGAN exhibited superior performance compared to their original counterparts across all metrics. While some metrics for MetricGAN also surpassed the original model, the improvements were less pronounced than those observed in CMGAN and MultiCMGAN.

\subsubsection{Ablation Study}

We conducted additional ablation studies using CMGAN on the VoiceBank+DEMAND dataset, with results presented in Table 3. When the SAFS was trained entirely in an unsupervised manner throughout the training process, the PESQ, CSIG, CBAK, COVL, and STOI metrics exhibited decreases of 0.083, 0.013, 0.137, 0.065, and 0.098, respectively. Similarly, omitting the pre-fine-tuning of DNSMOS for dynamic inputs led to a notable decline in most metrics, with the exception of CBAK. Additionally, removing the SNR-driven loss weighting resulted in degraded performance across metrics other than STOI. However, individual metrics alone do not provide a comprehensive assessment of the model's overall effectiveness, a point we will further elaborate on in the subsequent discussion of subjective evaluations.

\subsubsection{Subjective Effect Analysis}

To address the instability of certain metrics, such as STOI, we further analyzed the subjective enhancement effects of SaD. As illustrated in Figure 2, we randomly selected two distinct scenarios from the VoiceBank+DEMAND test set, with signal-to-noise ratios (SNRs) of 0.25 and 11.25 dB, respectively. The top row, from left to right, displays the results for the Noisy (SNR = 0.25 dB), CMGAN, and CMGAN+SaD conditions. The bottom row, from left to right, shows the results for the Noisy (SNR = 11.25 dB), CMGAN, and CMGAN+SaD conditions.

The CMGAN results were obtained using the official open-source weights and code for inference. Comparing the noisy and CMGAN results reveals that CMGAN tends to generate spurious high-frequency harmonics that do not exist in the original signal. In contrast, CMGAN+SaD effectively suppresses these pseudo-harmonics in both low-SNR and high-SNR scenarios, thereby improving the actual listening experience.

\begin{figure}[t]
  \begin{center}
  \includegraphics[width=7.5cm]{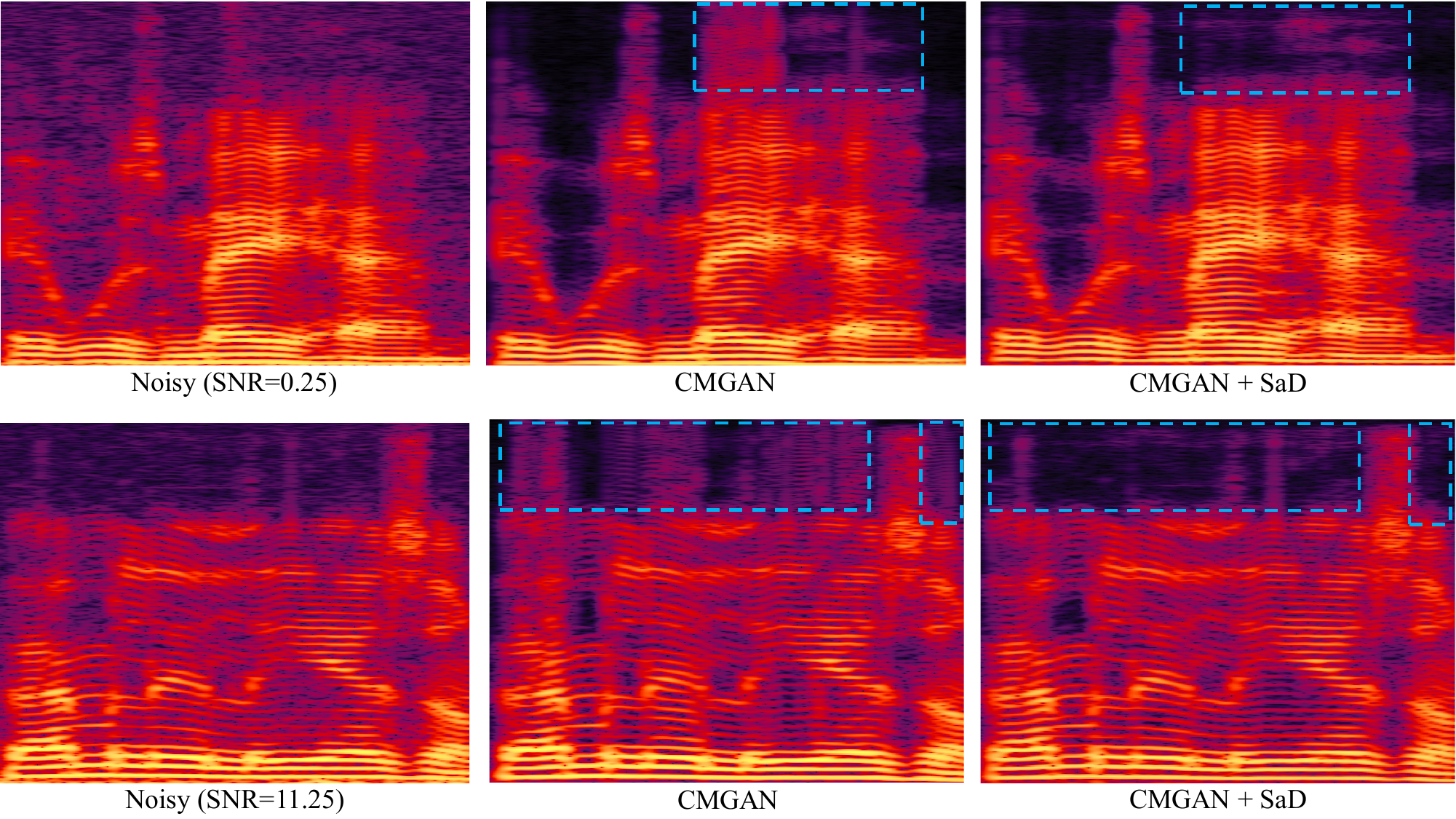}
  \caption{Frequency Analysis for Diverse Acoustic Scenarios. The upper and lower rows depict the time-frequency representations of two distinct scenarios. From left to right: the leftmost plot illustrates the input noisy signal, the middle plot shows the enhancement result obtained using the official CMGAN open-source code, and the rightmost plot presents the enhanced result achieved with CMGAN + SaD.}
  \label{fig:DFKD_overview}
  \end{center}
\end{figure}

\section{Conclusions}

This study introduces a scenario-aware discriminator tailored for SE models based on the GAN framework. Our approach integrates the time-frequency characteristics of the current acoustic scenario, partitions the enhanced speech generated by the generator into high- and low-frequency bands, and employs distinct quality evaluation metrics for each band. We further incorporate SNR information to distinguish the actual optimization objectives of the SE task. Based on this, we propose an SNR-driven method for adaptively adjusting loss weights, which effectively mitigates the need for extensive hyperparameter tuning during the training process of GAN-like models. Experimental results demonstrate that our method can further unlock the performance potential of various generator architectures, significantly enhancing their speech enhancement capabilities. The effectiveness of our approach is validated across multiple datasets and models, thereby confirming its feasibility and generalizability.

\newpage

\bibliographystyle{IEEEtran}
\bibliography{mybib}

\end{document}